# A joint-optimization NSAF algorithm based on the first-order Markov model

Yi Yu[1, 2], Haiquan Zhao[1, 2]

**Abstract:** Recently, the normalized subband adaptive filter (NSAF) algorithm has attracted much attention for handling the colored input signals. Based on the first-order Markov model of the optimal tap-weight vector, this paper provides a convergence analysis of the standard NSAF. Following the analysis, both the step size and the regularization parameter in the NSAF are jointly optimized in such a way that minimizes the mean square deviation. The resulting joint-optimization step size and regularization parameter (JOSR-NSAF) algorithm achieves a good tradeoff between fast convergence rate and low steady-state error. Simulation results in the context of acoustic echo cancellation demonstrate good features of the proposed algorithm.

## 1 Introduction

Adaptive filtering algorithms have been found in a wide range of practical applications such as system identification, channel equalization, beamforming, and echo cancellation [1]-[3]. Among these algorithms, a very popular algorithm is the normalized least mean square (NLMS), due to its low computational complexity and robust performance. Furthermore, to obtain fast convergence and low steady-state misadjustment (i.e., the final coefficient estimation error) simultaneously, many modified NLMS methods controlling the step size have been proposed, e.g., see [4]-[7] and references therein. However, these NLMS-type algorithms suffer from slow convergence when the input signals are colored, especially for the speech input signals.

To solve this problem, in a recent decade, the multiband-structure of the subband adaptive filter (SAF) has attracted much attention [3]. This is because the SAF divides the colored input signal into multiple mutually almost exclusive subband signals, and each decimated subband input signal is approximately white. What's more, as compared to the conventional subband structure, the multiband-structure has no band edge effects [3]. On the basis of this multiband-structure, the normalized SAF (NSAF) algorithm [8] was developed by Lee and Gan from the least perturbation principle. The NSAF exhibits faster convergence for the colored input signals than the NLMS, due mainly to the inherent decorrelating property of SAF [9]. Moreover, for high-order adaptive filter applications such as echo cancellation, the computational complexity of the NSAF is almost the same as that of the NLMS. It is worth mentioning that the NSAF is equivalent to the NLMS only when there is one subband. Afterwards, the theoretical models (including the transient and steady-state behavior) of the NSAF were provided in [10], [11]. Similar to the NLMS, the performance of the standard NSAF depends on two important parameters, i.e., the step size and the regularization parameter. The fixed step size governs a tradeoff between convergence rate and steady-state misadjustment. Specifically, for the NSAF, a large (small) step size leads to fast (slow) convergence rate but large (small) misadjustment in the steady-state. This conflict motivates the development of the NSAF with a variable step size (VSS) algorithms [12]-[17]. The original intention of the regularization parameter is to prevent the NSAF from numerical divergence when the $l_2$-norm of the input vector is very small or zero (this case is common in echo cancellation). Note that its value also reflects a compromise in the algorithm's performance like the step size does. Nevertheless, the only difference is that the directions of the step size and the regularization parameter controlling the algorithm's

Yi Yu
email: yuyi_xyuan@163.com

Haiquan Zhao
email: hqzhao_swjtu@126.com

1 Key Laboratory of Magnetic Suspension Technology and Maglev Vehicle, Ministry of Education, Southwest Jiaotong University, Chengdu, 610031, China.

2 School of Electrical Engineering at Southwest Jiaotong University, Chengdu, 610031, China.

performance are converse. Therefore, several variable regularization (VR) NSAF algorithms have also been proposed [18]-[21], which in a certain degree overcome the conflicting requirements of fast convergence rate and low misadjustment caused by the fixed regularization parameter. Although researchers have made some achievements on the optimization of these two parameters, many of the presented VSS-NSAF and VR-NSAF algorithms are essentially equivalent. Moreover, these algorithms are obtained based on the fact that one of two parameters is optimized by fixing the other.

In this paper, we firstly analyze the convergence performance of the standard NSAF based on the first-order Markov model of the optimal tap-weight vector. Second, a joint-optimization scheme of the step size and the regularization parameter is proposed by minimizing the mean square deviation (MSD) of the NSAF. The resulting algorithm is called the joint-optimization step size and regularization parameter NSAF (JOSR-NSAF) algorithm, which obtains improved performance.

## 2 Preliminary knowledge

Consider the observed data $d(n)$ that originates from the model

$$d(n) = \mathbf{u}^T(n)\mathbf{w}_o + \eta(n), \quad (1)$$

where $(\cdot)^T$ indicates the transpose, $\mathbf{w}_o$ is the unknown $M$-dimensional vector to be estimated with an adaptive filter, $\mathbf{u}(n) = [u(n), u(n-1), ..., u(n-M+1)]^T$ is the input signal vector, and $\eta(n)$ is the measurement noise which is assumed to be white Gaussian noise with zero-mean and variance $\sigma_\eta^2$. Fig. 1 shows the multiband-structure diagram of the SAF, where $N$ denotes number of subbands. The observed data $d(n)$ and input data $u(n)$ are partitioned into multiple subband signals $d_i(n)$ and $u_i(n)$ through the analysis filter bank, namely, $d_i(n) = d(n) * h_i$ and $u_i(n) = u(n) * h_i, i = 0,1,...,N-1$, where $h_i$ is the impulse response of the $i$th analysis filter $H_i(z)$ and $*$ denotes linear convolution. The subband output signals $y_i(n)$ are obtained by filtering the subband input signals $u_i(n)$ through an adaptive filter whose tap-weight vector is $\mathbf{w}(k) = [w_1(k), w_2(k), ..., w_M(k)]^T$. Then, the signals $y_i(n)$ and $d_i(n)$ are $N$-fold decimated [3], [8] to yield the signals $y_{i,D}(k)$ and $d_{i,D}(k)$ which are respectively formulated as

$y_{i,D}(k) = \mathbf{u}_i^T(k)\mathbf{w}(k-1)$ and $d_{i,D}(k) = d_i(kN)$, where $\mathbf{u}_i(k) = [u_i(kN), u_i(kN-1), ..., u_i(kN-M+1)]^T$. In this paper, we use $n$ to indicate the original sequences, and $k$ to indicate the decimated sequences. As shown in Fig. 2, the decimated subband error signals are expressed by subtracting $y_{i,D}(k)$ from $d_{i,D}(k)$ as

$$e_{i,D}(k) = d_{i,D}(k) - \mathbf{u}_i^T(k)\mathbf{w}(k-1), \quad i = 0,1,...,N-1. \quad (2)$$

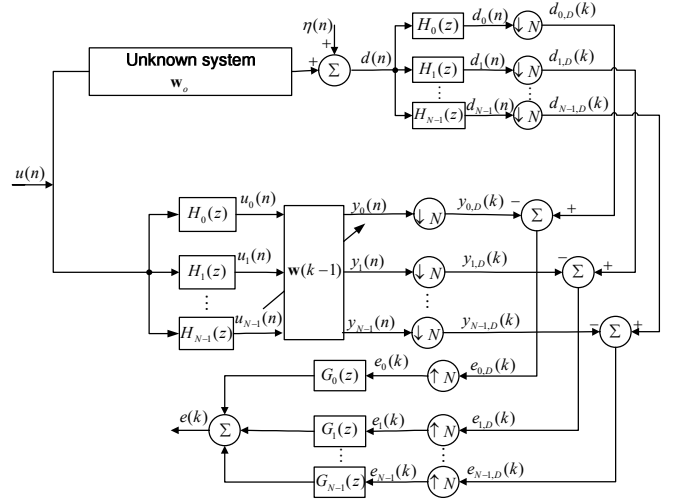

**Fig. 1.** Multiband-structure diagram of the SAF.

As reported in [8], the update equation of the standard NSAF algorithm is expressed as

$$\mathbf{w}(k) = \mathbf{w}(k-1) + \mu \sum_{i=0}^{N-1} \frac{e_{i,D}(k)\mathbf{u}_i(k)}{\delta + \|\mathbf{u}_i(k)\|^2} \quad (3)$$

where $\|\cdot\|$ denotes the $l_2$-norm of a vector, $\mu$ is the step-size, and $\delta > 0$ is a small regularization parameter.

## 3 Proposed JOSR-NSAF algorithm

In this section, the proposed JOSR-NSAF algorithm will be derived, whose inspiration comes from the joint-optimization NLMS (JO-NLMS) algorithm proposed by S. Ciochină *et al.* [7].

### 3.1 Some insights for convergence of the NSAF

Let us assume that the unknown vector $\mathbf{w}_o$ is a time-varying vector that follows a simplified first-order Markov model [24], i.e.,

$$\mathbf{w}_o(k) = \mathbf{w}_o(k-1) + \mathbf{q}(k) \quad (4)$$

where $\mathbf{q}(k)$ is a white Gaussian noise vector with zero-mean and covariance matrix $E[\mathbf{q}(k)\mathbf{q}^T(k)] = \sigma_q^2 \mathbf{I}_M$ with $\mathbf{I}_M$ being an $M \times M$ identity matrix and $E[\cdot]$ denoting the mathematical expectation. Evidently, the quantity $\sigma_q^2$ characterizes the randomness in $\mathbf{w}_o(k)$, and $\mathbf{q}(k)$ is independent of $\mathbf{w}_o(k-1)$.

Subtracting (4) from (3), we obtain

$$\tilde{\mathbf{w}}(k) = \tilde{\mathbf{w}}(k-1) - \mu \sum_{i=0}^{N-1} \frac{e_{i,\mathrm{D}}(k)\mathbf{u}_i(k)}{\delta + \|\mathbf{u}_i(k)\|^2} + \mathbf{q}(k) \quad (5)$$

where $\tilde{\mathbf{w}}(k) \triangleq \mathbf{w}_o(k) - \mathbf{w}(k)$ denotes the tap-weight error vector. Based on (1), (2) and (4), the decimated subband error signals can be rewritten as

$$e_{i,\mathrm{D}}(k) = \mathbf{u}_i^T(k)\tilde{\mathbf{w}}(k-1) + \mathbf{u}_i^T(k)\mathbf{q}(k) + \eta_i(k) \quad (6)$$

where $\eta_i(k)$ for $i = 0,1,\ldots,N-1$ are the subband noises that are obtained by partitioning the measurement noise $\eta(n)$, and have zero-mean and variances $\sigma_{\eta_i}^2 = \sigma_\eta^2 / N$ [11], [22].

Taking the squared $l_2$-norm and mathematical expectation on both sides of (6), and removing the uncorrelated product of $\mathbf{q}(k)$ and $\tilde{\mathbf{w}}(k-1)$, we get

$$\mathrm{MSD}(k) = \mathrm{MSD}(k-1) + M\sigma_q^2$$
$$-2\mu \sum_{i=0}^{N-1} E\left[\frac{e_{i,\mathrm{D}}(k)\tilde{\mathbf{w}}^T(k-1)\mathbf{u}_i(k)}{\delta + \|\mathbf{u}_i(k)\|^2}\right]$$
$$-2\mu \sum_{i=0}^{N-1} E\left[\frac{e_{i,\mathrm{D}}(k)\mathbf{q}^T(k)\mathbf{u}_i(k)}{\delta + \|\mathbf{u}_i(k)\|^2}\right] \quad (7)$$
$$+\mu^2 \sum_{i=0}^{N-1} E\left[\frac{e_{i,\mathrm{D}}^2(k)\mathbf{u}_i^T(k)\mathbf{u}_i(k)}{(\delta + \|\mathbf{u}_i(k)\|^2)^2}\right]$$

where $\mathrm{MSD}(k) \triangleq E[\|\tilde{\mathbf{w}}(k)\|^2]$ denotes the MSD of the algorithm at the $k$th iteration. In (7), we also use the diagonal assumption, i.e., $E[\mathbf{u}_i^T(k)\mathbf{u}_j(k)] \approx 0$, $i \neq j$, which has been used in the derivation of the standard NSAF [8]. For a long adaptive filter, it is assumed that the fluctuation of $\|\mathbf{u}_i(k)\|^2$ from one iteration to the next is small enough [12], [16] so that (7) becomes

$$\mathrm{MSD}(k) = \mathrm{MSD}(k-1) + M\sigma_q^2$$
$$-2\mu \sum_{i=0}^{N-1} \frac{E[e_{i,\mathrm{D}}(k)\tilde{\mathbf{w}}^T(k-1)\mathbf{u}_i(k)]}{E[\delta + \|\mathbf{u}_i(k)\|^2]}$$
$$-2\mu \sum_{i=0}^{N-1} \frac{E[e_{i,\mathrm{D}}(k)\mathbf{q}^T(k)\mathbf{u}_i(k)]}{E[\delta + \|\mathbf{u}_i(k)\|^2]} \quad (8)$$
$$+\mu^2 \sum_{i=0}^{N-1} \frac{E[e_{i,\mathrm{D}}^2(k)\mathbf{u}_i^T(k)\mathbf{u}_i(k)]}{E[(\delta + \|\mathbf{u}_i(k)\|^2)^2]}$$

Owing to the inherent decorrelating property of SAF, we can assume that each decimated subband input signal is close to a white signal, i.e., $\mathbf{u}_i(k)\mathbf{u}_i^T(k) \approx \mathbf{I}_M \sigma_{u_i}^2(k)$ and $\mathbf{u}_i^T(k)\mathbf{u}_i(k) \approx M\sigma_{u_i}^2(k)$ [14]. Hence, (8) is changed as

$$\mathrm{MSD}(k) = \mathrm{MSD}(k-1) + M\sigma_q^2$$
$$-2\mu \sum_{i=0}^{N-1} \frac{E[e_{i,\mathrm{D}}(k)\tilde{\mathbf{w}}^T(k-1)\mathbf{u}_i(k)]}{\delta + M\sigma_{u_i}^2(k)}$$
$$-2\mu \sum_{i=0}^{N-1} \frac{E[e_{i,\mathrm{D}}(k)\mathbf{q}^T(k)\mathbf{u}_i(k)]}{\delta + M\sigma_{u_i}^2(k)} \quad (9)$$
$$+\mu^2 \sum_{i=0}^{N-1} \frac{E[e_{i,\mathrm{D}}^2(k)\mathbf{u}_i^T(k)\mathbf{u}_i(k)]}{(\delta + M\sigma_{u_i}^2(k))^2}$$

To further proceed, the commonly used independent assumption [1], [7], [10], [22] that $\tilde{\mathbf{w}}(k-1)$, $\mathbf{u}_i(k)$, $\mathbf{q}(k)$ and $\eta_i(k)$ are statistically independent is necessary. With this assumption and using the Gaussian moment factoring theorem [1], [7], and after some manipulations, we have

$$E[e_{i,\mathrm{D}}(k)\tilde{\mathbf{w}}^T(k-1)\mathbf{u}_i(k)] \approx \sigma_{u_i}^2(k)\mathrm{MSD}(k-1), \quad (10)$$

$$E[e_{i,\mathrm{D}}(k)\mathbf{q}^T(k)\mathbf{u}_i(k)] \approx M\sigma_q^2 \sigma_{u_i}^2(k), \quad (11)$$

$$E[e_{i,\mathrm{D}}^2(k)\mathbf{u}_i^T(k)\mathbf{u}_i(k)] \approx M\sigma_{u_i}^2(k)\sigma_{\eta_i}^2$$
$$+(M+2)\sigma_{u_i}^4(k)[\mathrm{MSD}(k-1) + M\sigma_q^2]. \quad (12)$$

Substituting (10)-(12) into (9), then (9) becomes

$$\mathrm{MSD}(k) = \hbar(\mu,\delta)\mathrm{MSD}(k-1) + \varphi(\mu,\delta) \quad (13)$$

where

$$\hbar(\mu,\delta) =$$
$$\left[1 - 2\mu \sum_{i=0}^{N-1} \frac{\sigma_{u_i}^2(k)}{\delta + M\sigma_{u_i}^2(k)} + \mu^2 \sum_{i=0}^{N-1} \frac{(M+2)\sigma_{u_i}^4(k)}{(\delta + M\sigma_{u_i}^2(k))^2}\right], \quad (14)$$

$$\varphi(\mu,\delta) = \hbar(\mu,\delta)M\sigma_q^2 + \mu^2 \sum_{i=0}^{N-1} \frac{M\sigma_{u_i}^2(k)\sigma_{\eta_i}^2}{(\delta + M\sigma_{u_i}^2(k))^2}. \quad (15)$$

Evidently, the relation (13) consists of two parts, i.e., $\hbar(\mu,\delta)$ and $\varphi(\mu,\delta)$, which reveal the convergence and misadjustment behavior of the NSAF, respectively.

**Remark 1**: The term $\hbar(\mu,\delta)$ controls the convergence rate of the algorithm in mean square sense. As can be seen, the convergence rate is dependent on the step size, regularization parameter, filter length, number of subbands, and subband input variances. Interestingly, the convergence rate is not influenced by the subband noise variances $\sigma_{\eta_i}^2$ and the model uncertainties $\sigma_q^2$. In addition, some classical convergence conclusions can be obtained by analyzing the convergence term $\hbar(\mu,\delta)$:

1) The fastest convergence rate of the algorithm is obtained when the value of $\hbar(\mu, \delta)$ is minimum. Therefore, setting the derivative of $\hbar(\mu, \delta)$ with respect to the step size to zero, the optimal step size for ensuring the fastest convergence rate is obtained as

$$\mu_{\text{opt-con}} = \frac{\sum_{i=0}^{N-1}\sigma_{u_i}^2(k)\left[\delta + M\sigma_{u_i}^2(k)\right]/\left[\delta + M\sigma_{u_i}^2(k)\right]^2}{\sum_{i=0}^{N-1}(M+2)\sigma_{u_i}^4(k)/\left[\delta + M\sigma_{u_i}^2(k)\right]^2}. \quad (16)$$

After neglecting the regularization parameter (i.e., $\delta = 0$) and supposing a long filter (i.e., $M \gg 2$), (16) can be approximated as $\mu_{\text{opt-con}} \approx 1$, which is a well-known result for the standard NSAF[3].

2) To ensure the mean square stability of the NSAF algorithm, the range of the step size can be formulated by imposing $|\hbar(\mu, \delta)| < 1$ as

$$0 < \mu_{\text{stability}} < 2\mu_{\text{opt-con}}. \quad (17)$$

By again taking $\delta = 0$ and $M \gg 2$, we obtain the stability range presented in [3], [8], i.e., $0 < \mu_{\text{stability}} < 2$.

**Remark 2**: The term $\varphi(\mu, \delta)$ in (13) determines the misadjustment of the NSAF algorithm. Evidently, the misadjustment depends on $\sigma_q^2$ and $\sigma_{\eta_i}^2$, and increases as these two quantities increase. It is worth to note that the smallest misadjustment of the algorithm can be obtained by the minimization of $\varphi(\mu, \delta)$. As a consequence, by setting the derivative of $\varphi(\mu, \delta)$ with respect to the step size to zero, the optimal step size for obtaining the smallest misadjustment is expressed as

$$\mu_{\text{opt-mis}} = \frac{\sigma_q^2 \sum_{i=0}^{N-1}\sigma_{u_i}^2(k)\left[\delta + M\sigma_{u_i}^2(k)\right]/\left[\delta + M\sigma_{u_i}^2(k)\right]^2}{\sum_{i=0}^{N-1}\left[(M+2)\sigma_q^2\sigma_{u_i}^4(k) + \sigma_{u_i}^2(k)\sigma_{\eta_i}^2(k)\right]/\left[\delta + M\sigma_{u_i}^2(k)\right]^2}. \quad (18)$$

Assuming that the unknown system is stationary, i.e., $\sigma_q^2 \approx 0$, (18) will lead to $\mu_{\text{opt-mis}} \approx 0$. This result implies that the step size should be very small (e.g., close to zero) to obtain small misadjustment.

**Remark 3**: From Remarks 1 and 2, it is concluded that the fixed step size determines the convergence rate and misadjustment of the NSAF algorithm in opposite directions. In other words, using the fixed step size is unrealistic to obtain the NSAF's desired performance including both fast convergence rate and small misadjustment. Hence, this conclusion motivates the VSS methods to meet these two performances. In all the VSS schemes, there is a common fact that the step size gradually decreases as the algorithm converges from the starting stage to the steady-state stage. Although the regularization constant in (3) is originally introduced to avoid the numerical instability of the NSAF when the $l_2$-norm of the subband input signals is very small (in extreme case, is zero), its value also influences the convergence rate and misadjustment of the algorithm [20]. Interestingly, the influence of the regularization constant on these two performances is opposite to that of the step size. That is to say, as the regularization constant increases, the convergence rate will become slow while the misadjustment will decrease. As a result, a potential scheme is to control these two parameters simultaneously to improve the performance of the NSAF, which will be described in following subsection.

### 3.2 A joint-optimization scheme

Using a time-varying step size $\mu_i(k)$ and a time-varying regularization parameter $\delta_i(k)$ for $i = 0,1,...,N-1$, (13) can be rewritten as

$$\text{MSD}(k) = \left\{1 - 2\sum_{i=0}^{N-1}\frac{\mu_i(k)\sigma_{u_i}^2(k)}{\delta_i(k) + M\sigma_{u_i}^2(k)} + \sum_{i=0}^{N-1}\frac{\mu_i^2(k)(M+2)\sigma_{u_i}^4(k)}{\left[\delta_i(k) + M\sigma_{u_i}^2(k)\right]^2}\right\} \times \left[\text{MSD}(k-1) + M\sigma_q^2\right] + \sum_{i=0}^{N-1}\frac{\mu_i^2(k)M\sigma_{u_i}^2(k)\sigma_{\eta_i}^2}{\left[\delta + M\sigma_{u_i}^2(k)\right]^2}. \quad (19)$$

In order to minimize the MSD of the NSAF at each iteration, the following subband constraints are imposed, i.e.,

$$\frac{\partial \text{MSD}(k)}{\partial \mu_i(k)} = 0 \text{ and } \frac{\partial \text{MSD}(k)}{\partial \delta_i(k)} = 0, \ i = 0,1,...,N-1. \quad (20)$$

Applying (20), a joint-optimization strategy of $\mu_i(k)$ and $\delta_i(k)$ for each subband is obtained as,

$$\frac{\mu_i(k)}{\delta_i(k) + M\sigma_{u_i}^2(k)} = \frac{\text{MSD}(k-1) + M\sigma_q^2}{(M+2)\sigma_{u_i}^2(k)\left[\text{MSD}(k-1) + M\sigma_q^2\right] + M\sigma_{\eta_i}^2}. \quad (21)$$

Substituting (21) into (3), we obtain a new tap-weight update expression

$$\mathbf{w}(k) = \mathbf{w}(k-1)$$
$$+ \sum_{i=0}^{N-1} \frac{\left[\text{MSD}(k-1) + M\sigma_q^2\right] e_{i,\text{D}}(k)\mathbf{u}_i(k)}{(M+2)\sigma_{u_i}^2(k)\left[\text{MSD}(k-1) + M\sigma_q^2\right] + M\sigma_{\eta_i}^2} \quad . (22)$$

Likewise, substituting (21) into (19) and after some simple computations, the parameter $\text{MSD}(k-1)$ in (22) is updated by

$$\text{MSD}(k) = \left[\text{MSD}(k-1) + M\sigma_q^2\right] \times$$
$$\left\{1 - \sum_{i=0}^{N-1} \frac{\left[\text{MSD}(k-1) + M\sigma_q^2\right]\sigma_{u_i}^2(k)}{(M+2)\sigma_{u_i}^2(k)\left[\text{MSD}(k-1) + M\sigma_q^2\right] + M\sigma_{\eta_i}^2}\right\} .(23)$$

### 3.3 Convergence of the proposed algorithm

Let us define the decimated *a priori* error of the *i*th subbands as $e_{a,i}(k) \triangleq \tilde{\mathbf{w}}^T(k-1)\mathbf{u}_i(k)$, we have $E\left[e_{a,i}^2(k)\right] = \text{MSD}(k-1)\sigma_{u_i}^2(k)$, then (23) can be changed as

$$\text{MSD}(k) = \beta(k)\text{MSD}(k-1) + \beta(k)M\sigma_q^2 \quad (24)$$

where

$$\beta(k) =$$
$$1 - \sum_{i=0}^{N-1} \frac{E\left[e_{a,i}^2(k)\right] + M\sigma_q^2\sigma_{u_i}^2(k)}{(M+2)\left[E\left[e_{a,i}^2(k)\right] + M\sigma_q^2\sigma_{u_i}^2(k)\right] + M\sigma_{\eta_i}^2} \quad . (25)$$

By continuously iterating (24), we get

$$\text{MSD}(k) = \left(\beta(k)\cdots\beta(2)\beta(1)\right)\text{MSD}(0)$$
$$+ M\sigma_q^2 \sum_{j=1}^{k}\beta(j)\cdots\beta(k) \quad , (26)$$

$$\text{MSD}(k-1) = \left(\beta(k-1)\cdots\beta(2)\beta(1)\right)\text{MSD}(0)$$
$$+ M\sigma_q^2 \sum_{j=1}^{k-1}\beta(j)\cdots\beta(k-1) \quad . (27)$$

Combining (26) and (27), a relation is founded as
$$\Delta\text{MSD}(k) = \text{MSD}(k) - \text{MSD}(k-1) =$$
$$\left(\beta(k)-1\right)\left(\beta(k-1)\cdots\beta(2)\beta(1)\right)\text{MSD}(0) + M\sigma_q^2\beta(k) \quad . (28)$$

Again using the assumption of a long adaptive filter, i.e., $M \gg 2$, which is the property of echo cancellation application (e.g., $M = 512$ in the following simulation section), thus from (25) we obtain

$$\beta(k) \approx 1 - \sum_{i=0}^{N-1} \frac{E\left[e_{a,i}^2(k)\right] + M\sigma_q^2\sigma_{u_i}^2(k)}{M\left(E\left[e_{a,i}^2(k)\right] + M\sigma_q^2\sigma_{u_i}^2(k)\right) + M\sigma_{\eta_i}^2} \quad (29)$$
$$< \left(1 - \frac{N}{M}\right) = \beta_{\max} < 1$$

To ensure the mean square stability of the proposed algorithm, the MSD must decrease iteratively, i.e., $\Delta\text{MSD}(k) < 0$. Thus, the quantity $\sigma_q^2$ has to satisfy the inequality

$$\sigma_q^2 < \frac{(1-\beta(k))}{M\beta(k)}\beta_{\max}^{k-1}\text{MSD}(0) . \quad (30)$$

Under the condition of (28) the algorithm has reached steady-state, and then the following relation holds

$$\text{MSD}(\infty) =$$
$$\lim_{k\to\infty}\left(\beta(k)\cdots\beta(2)\beta(1)\text{MSD}(0) + M\sigma_q^2\sum_{j=1}^{k}\beta(j)\cdots\beta(k)\right)$$
$$< \lim_{k\to\infty}\left(\beta_{\max}^k \text{MSD}(0) + M\sigma_q^2 \frac{\beta_{\max} - \beta_{\max}^{k+1}}{1-\beta_{\max}}\right) \quad .(31)$$
$$= M\sigma_q^2 \frac{\beta_{\max}}{1-\beta_{\max}}$$

The formula (31) reveals that the convergence of the proposed JOSR-NSAF is stable in mean square sense.

### 3.4 Practical considerations

To implement the above-presented JOSR-NSAF algorithm, some practical considerations about parameters $\sigma_{u_i}^2(k)$, $\sigma_\eta^2$, and $\sigma_q^2$ are necessary which are listed below.

1) The subband input variances $\sigma_{u_i}^2(k)$ for $i = 0,1,...,N-1$ can be estimated by $\hat{\sigma}_{u_i}^2(k) = \mathbf{u}_i^T(k)\mathbf{u}_i(k)/M$ [7], [23].

2) The second consideration is to take the measurement noise variance $\sigma_\eta^2$, which also appears in many VSS and VR versions of the NSAF algorithm, e.g., [12], [13], [15], [16], [19], [21]. Usually, in practical applications, $\sigma_\eta^2$ can be easily estimated. Also, several different methods based on an exponential window have been developed to estimate this variance [4], [5], [12]. For example, in echo cancellation, it can be estimated during silences of the near-end talker, i.e., in a single-talk scenario [12]. Importantly, discussing the performance of these methods estimating $\sigma_\eta^2$ is not the purpose of this work.

3) The only remaining consideration is how to choose $\sigma_q^2$, which plays a very important role in the performance of the proposed JOSR-NSAF. For a small $\sigma_q^2$, the algorithm has a small steady-state misadjustment but a poor tracking capability; conversely, a large $\sigma_q^2$ results in a good tracking performance but increases the steady-state misadjustment. To solve this compromise, therefore, $\sigma_q^2$ is estimated as [7], [23]

$$\hat{\sigma}_q^2(k) = \|\mathbf{w}(k) - \mathbf{w}(k-1)\|^2 / M. \qquad (32)$$

This relation is obtained by taking the $l_2$-norm on both sides of (4) and replacing $\mathbf{w}_o(k)$ by its estimate $\mathbf{w}(k)$. As can be seen, in the initial stage of adaptation or when the unknown system suddenly changes, the value of $\hat{\sigma}_q^2$ is large, thus leading to fast convergence rate and good tracking capability. Moreover, when the algorithm goes into the steady-state, the value of $\hat{\sigma}_q^2$ is small, thus obtaining low steady-state misadjustment.

Based on the above considerations, the proposed JOSR-NSAF algorithm is summarized in Table 1. Note that, the JOSR-NSAF will reduce to the JO-NLMS in [7] when the number of subbands is one.

**Table 1** Summary of the proposed JOSR-NSAF algorithm

| Initializations | $\mathbf{w}(0) = \mathbf{0}$, $\mathrm{MSD}(0) = 1$ |
|---|---|
| Parameters | $\sigma_\eta^2$, noise variance known or estimated |
| Adaptive process | for $i = 0, 1, ..., N-1$ <br> $e_{i,D}(k) = d_{i,D}(k) - \mathbf{u}_i^T(k)\mathbf{w}(k-1)$ <br> $\hat{\sigma}_{u_i}^2(k) = \mathbf{u}_i^T(k)\mathbf{u}_i(k)/M$ <br> end <br> $g(k) = \mathrm{MSD}(k-1) + M\hat{\sigma}_q^2(k-1)$ <br> $\pi_i(k) = \dfrac{g(k)}{(M+2)\sigma_{u_i}^2(k)g(k) + M\sigma_{\eta_i}^2}$ <br> $\mathbf{w}(k) = \mathbf{w}(k-1) + \sum_{i=0}^{N-1}\pi_i(k)e_{i,D}(k)\mathbf{u}_i(k)$ <br> $\mathrm{MSD}(k) = \left[1 - \sum_{i=1}^{N-1}\pi_i(k)\sigma_{u_i}^2(k)\right]g(k)$ <br> $M\hat{\sigma}_q^2(k) = \|\mathbf{w}(k) - \mathbf{w}(k-1)\|^2$ |

## 4 Simulation results

To evaluate the performance of the algorithm, extensive simulations are performed in the context of acoustic echo cancellation. In our simulations, the unknown vector $\mathbf{w}_o$ to be identified is a room acoustic echo path with $M = 512$ taps. Also, to show the tracking capability of the algorithm, the unknown vector is changed abruptly from $\mathbf{w}_o$ to $-\mathbf{w}_o$ in the middle of input samples. The colored input signal is either an AR(1) process with a pole at 0.95 or a speech signal. The measurement noise $\eta(n)$ is white Gaussian with a signal-to-noise ratio (SNR) of either 30 dB or 20 dB.

It is assumed that the variance of the measurement noise, $\sigma_\eta^2$, is known, because it can be easily estimated like the ways in [4, 5, 12]. A cosine modulated filter bank [3] is used in all the SAF algorithms. As a measure of the algorithm performance, the normalized MSD (NMSD) (or called the misadjustment) is defined as $10 \times \log_{10}(\|\mathbf{w}_o - \mathbf{w}(k)\|_2^2 / \|\mathbf{w}_o\|_2^2)$ (dB). All results are obtained by averaging over 30 independent runs, except for speech input (single realization).

We first examine the performance of the JO-NLMS in [7] and proposed JOSR-NSAF (with $N = 2$ and 8 subbands) algorithms for an AR(1) input, as shown in Fig. 2. From this figure, it can be noted that the JOSR-NSAF algorithm has faster convergence rate than the JO-NLMS (i.e., the JOSR-NSAF with $N = 1$) algorithm for the colored input signal. Moreover, with an increased number of subbands $N$, the convergence rate is further improved. The reason behind this phenomenon is that each decimated subband input signal is closer to a white signal for a larger number of subbands. In the following simulations, we choose $N = 8$ to compare all the NSAF-type algorithms.

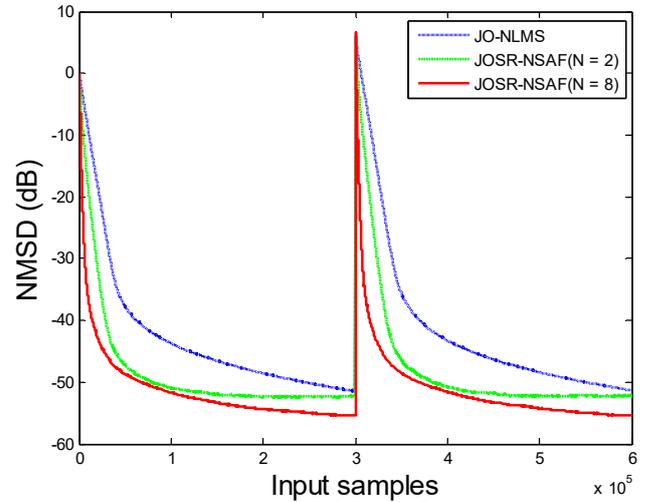

**Fig. 2.** NMSD curves of the JO-NLMS and proposed JOSR-NSAF (with $N = 2$ and 8) algorithms. $\mathrm{SNR} = 30\,\mathrm{dB}$, AR(1) input.

Then, Fig. 3 shows the NMSD performance of the standard NSAF (with $\mu = 1$ and 0.05), VSSM-NSAF [12], NVSS-NSAF [16], VRM-NSAF [19], and proposed JOSR-NSAF algorithms using an AR(1) process as input signal. All these VSS and VR algorithms require the priori knowledge of the measurement noise variance $\sigma_\eta^2$, so we assume that its value is available to obtain a fair comparison. Also, we set the algorithms' parameters according to the recommendations provided in [12], [16], [19]. As can be seen, compared with the NSAF algorithm, its VSS and VR versions improve the performance in terms of the convergence rate and steady-state misadjustment. Importantly, the improvement of the proposed JOSR-NSAF in the steady-state performance is more obvious than its counterparts. It can also be observed from Fig. 3 that as the SNR decreases (or the measurement noise variance $\sigma_\eta^2$ increases), the steady-state misadjustment of these NSAFs increases but the convergence rate of that does not change, which is consistent with the previous analysis results in Remarks 1 and 2.

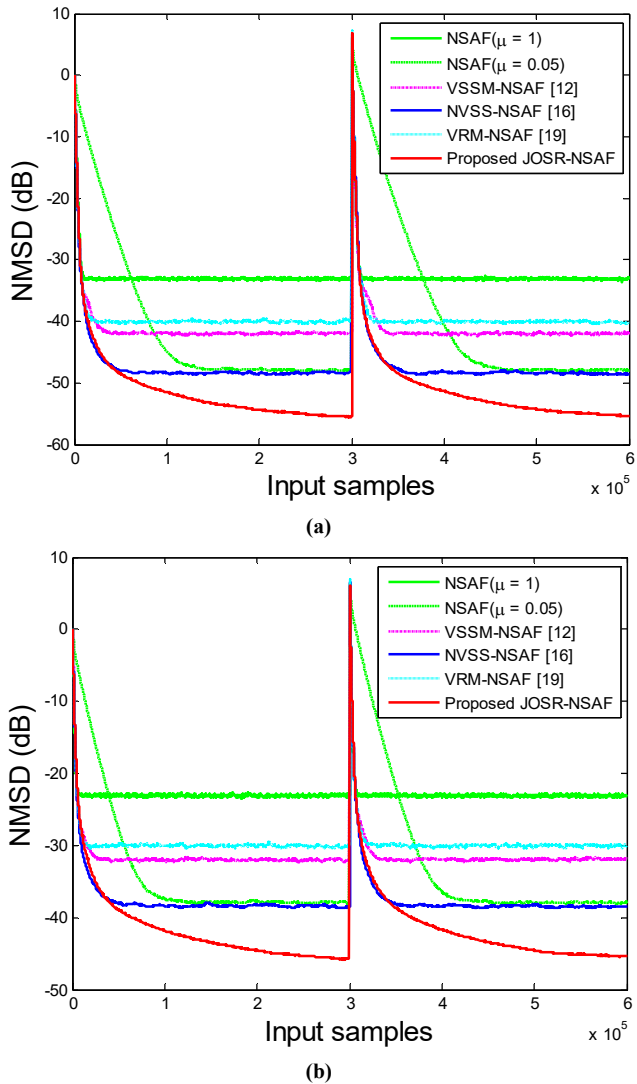

**Fig. 3.** NMSD curves of various NSAF-type algorithms for AR(1) input signal. (a) $\text{SNR} = 30 \text{ dB}$ ; (b) $\text{SNR} = 20 \text{ dB}$ . VSSM-NSAF: $\kappa = 6$ ; NVSS-NSAF: $\kappa = 3, \lambda = 4$ ; VRM-NSAF: $\alpha = 0.995, Q = 1000$ . The regularization parameter for the NSAF, VSSM-NSAF and NVSS-NSAF algorithms is chosen as $\delta = 10\sigma_{u_i}^2$ .

Finally, Fig. 4 compares the performance of the proposed JOSR-NSAF with that of the NSAF (with $\mu = 1$ ), VSSM-NSAF, NVSS-NSAF, and VRM-NSAF in speech input scenario. These results are similar to those results with AR(1) input in Fig. 3, which demonstrates that the proposed algorithm also works better than the existing VSS and VR NSAF algorithms for speech input signal. In addition, the proposed algorithm does not require any additional parameters to control its performance relative to many of its counterparts.

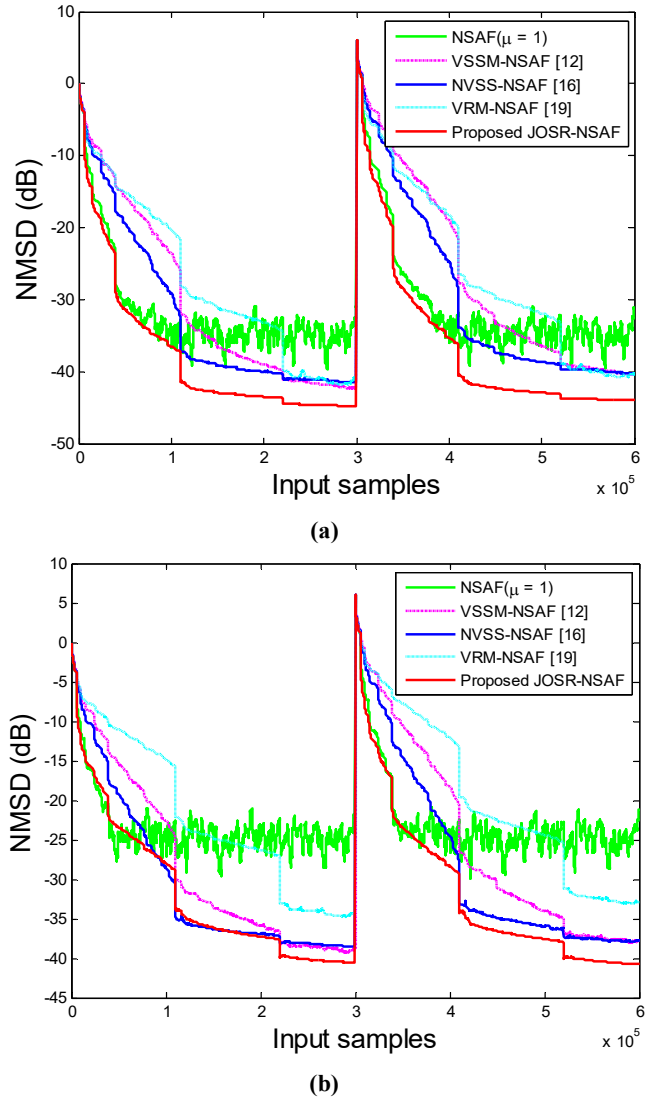

**Fig. 4.** NMSD curves of various NSAF-type algorithms for speech input signal. The choice of the algorithms' parameters is the same as Fig. 3. (a) $\text{SNR} = 30 \text{ dB}$ ; (b) $\text{SNR} = 20 \text{ dB}$ .

## 5 Conclusions

We have analyzed the convergence performance of the standard NSAF using a first-order Markov model of the optimal tap-weight vector. Based on this model, a joint-optimization NSAF algorithm has been derived by minimizing the MSD of the NSAF over both the step size and the regularization parameter, aiming to simultaneously obtain fast convergence rate and low steady-state misadjustment. Simulation results in acoustic echo cancellation application have shown that the proposed algorithm outperforms many existing VSS and VR versions of the NSAF in performance.

**Acknowledgment**

This work was partially supported by National Science Foundation of P.R. China (Grant: 61271340, 61571374 and 61433011), and the Fundamental Research Funds for the Central Universities (Grant: SWJTU12CX026).